\documentclass[aps,prl,twocolumn]{revtex4}
\usepackage{amsmath}    
\usepackage{graphicx}   
\usepackage{verbatim}   
\usepackage{color}      
\usepackage{subfigure}  
\usepackage{hyperref}   
\usepackage{amssymb}
\usepackage{dcolumn}
\usepackage{bm}

\begin{document}

\title{Density profile of multi-state fuzzy dark matter}%

\author{Lauren Street}
\email{streetlg@mail.uc.edu}
\affiliation{Department of Physics, University of Cincinnati}

\author{Peter Suranyi}
\email{peter.suranyi@gmail.com}
\affiliation{Department of Physics, University of Cincinnati}

\author{L.C.R. Wijewardhana}
\email{rohana.wijewardhana@gmail.com}
\affiliation{Department of Physics, University of Cincinnati}

\date{\today}

\begin{abstract}
Equations of motion for excited states of weakly self-interacting bosons forming fuzzy dark matter are solved using the WKB approximation. The contribution of self-interactions are neglected in the equations of motion. Wave functions of excited states are expressed in terms of a yet undetermined gravitational potential. At equilibrium, the contributions of states to the density distribution are summed using Bose-Einstein statistics. Combined with the Poisson equation, a differential equation is obtained for the gravitational potential, which has physically acceptable solutions only if the energy spectrum of excited states has a finite gap, corresponding to a finite virial radius. Such a gap could be created by decay processes, in first order perturbation of the self-interaction potential.  The obtained  density profile is found to be similar to the Burkert profile.
\end{abstract}

\maketitle

\section{Introduction}
\noindent The structure of galaxies and the rotation curves of stars in galaxies cannot be explained without the assumption that most of galactic matter is composed of presently unknown particles, termed dark matter (DM), which interact very weakly with particles of the Standard Model.  One of the most popular variants of DM is the WIMP, consisting of massive, non-relativistic particles, heavier than neutrinos~\cite{Peebles,Bond,Blumenthal}. Since no such particles, in the appropriate mass range, have been discovered yet, other alternatives for DM have also been considered. Among others, prominent candidates are ultralight bosons, with Compton wavelengths of cosmic size~\cite{Ruffini,Sin,Hu,BH,Hui}.

Simulations of collapsing systems of ultralight bosons, interacting only through gravity, were performed recently by~\cite{Schive, Schive2, Schwabe, Veltmaat, Du,Levkov_Panin}.  Bosonic systems were shown to collapse to a condensed core, surrounded by a virialized halo of non-relativistic bosons. In a subsequent work~\cite{Lin}, numerical solutions of excited states of the Schr\"odinger-Poisson (SP) equations were calculated self-consistently.  The authors showed that there is a viable description of galactic DM consisting of a condensed core surrounded by a halo composed of excited eigenstates. The relative weight of excited states in the system was fixed, using several parameters, including the effective inverse temperature, $\beta$, chemical potential, $\mu$, cutoff in the binding energy, $E_c$, or other scale parameters. Recently, simulations have also been performed for ultralight bosons with self-interactions \cite{Glennon_Chanda}.  

The purpose of this work is to construct DM from self-adjoint or complex ultralight bosons with attractive or repulsive self-interactions.  We ignore self-interactions in solving the equations of motion, but consider the effect of $2 \rightarrow 2$ interactions on the stability of excited eigenstates.  We emphasize that,  because only this particular interaction is relevant, our model can be used for real, or just as easily for complex, scalar fields. For the sake of simplicity, we focus on a real scalar field giving rise to an axion-like particle (ALP) subject to a $\Phi^4$ self-interaction.  

For the  range of the total mass of DM in a galaxy considered here, the contribution of self-interaction terms to the equations of motion is negligible compared to that of the gravitational interaction. The ratio of self-interaction to gravitational interactions scales as
\begin{equation*}
\frac{\text SI}{GI}\sim \frac{M_P^2}{f_a^2}\frac{1}{m_a^2r_s^2},   
\end{equation*}
where $M_P = G^{-1/2}$ is the Planck mass and $G$ is Newton's constant, $f_a$ is the axion decay constant, $r_s$ is the radial scale of the system and $m_a$ is the mass of the axion.

Self-interactions of bosons may be important for extremely large galaxies. In fact, based on studies of axion stars~\cite{ChavanisMR,ChavanisMR2,Eby1} they can possibly generate a cutoff in the mass spectrum of stable, extremely large galaxies with very large densities. That possibility will be investigated in  future publications. 

\section{WKB approximation to excited states}
The main purpose of this letter is to find an equation for the gravitational potential, $V_g$, of an FDM halo.  Therefore, radial wave functions, $\psi_{nl}$, and the total density of dark matter, $\rho$, will be calculated as functions of the yet unknown $V_g$. Radial wave functions of excited states, labeled by principal quantum number $n$ and angular quantum number $l$,  satisfy the Schr\"odinger (Gross-Pitaevskii) equation
\begin{align}\label{scheq} 
E_{nl}\psi_{nl}=-\frac{1}{2m_a}\left[\psi_{nl}''+\frac{2}{r}\psi_{nl}'\right]+\left[\frac{1}{2 m_a}\frac{l(l+1)}{r^2}+V_g\right]\psi_{nl},
\end{align}
where, assuming spherical symmetry, the gravitational potential $V_g$ is given by
\begin{align*}
V_g(r)=- G \, m_a\, \int d^3r'\frac{\rho(r')}{|\vec r-\vec r'|}.
\end{align*}
and where
\begin{align*}
    \rho(r)=m_a\sum_{nl}(2l+1)N_{nl}|\psi_{nl}|^2,
\end{align*}
with $N_{nl}$ representing the occupation number of  states with quantum numbers $n$ and $l$.  Notice that the factor $2l+1$ arises due to taking a sum over the magnetic quantum number.

Normalizing wave functions as
\begin{align*}
    \int d^3r\,|\psi_{nl}|^2=1
\end{align*}
implies that
\begin{align*}
    \int d^3r\rho(r)=m_a\sum_{nl}(2l+1)N_{nl}=M= m_a\,N,
\end{align*}
where $N$ is the total number of axions. For a self-adjoint scalar, like an axion, $N$ is not conserved. However, as it has been shown in~\cite{ESW,EMSW,Braaten}, the decrease of $N$, due to the decay of bound axions into relativistic axions or other elementary particles, is negligible during the lifetime of the universe for so called ``dilute" axion stars.  However, it has recently been shown that this assumption is only valid for particles with relatively small decay constants compared to the Planck scale~\cite{ESSW}

The dimensionless and positive scaling function, $v(z)$, of the gravitational potential and scaled radial coordinate, $z=r/r_s$ are introduced as
\begin{align}\label{gravpot}
   - V_g / G =\frac{m_a^2}{r_s}\int d^3z'\frac{\tilde\rho(z')}{|z-z'|}=\frac{M\, m_a}{r_s}v(z),
\end{align}
where $\tilde\rho(z) = \left(r_s^3/m_a\right) \rho(r)$ is the rescaled density.

 (\ref{gravpot}) defines only the ratio $v(z)/r_s$, not $r_s$ and $v(z)$ separately.  In what follows, we choose a ``natural" definition,
\begin{align}\label{rsdef}
\frac{1}{r_s} = \left\langle \frac{1}{r} \right\rangle,
\end{align}
which also implies $v(0) = 1$.  Note that phenomenological models have definite choices for $r_s$ which may differ by finite, $\mathcal{O}(1)$ factors from the value defined by (\ref{rsdef}). 

We define the dimensionless size parameter, $S$, for the galactic fuzzy dark matter (FDM) as
\begin{align*}
    S=2\, G M m_a^2 r_s \gg 1.
\end{align*}
We also define the dimensionless rescaled parameters,
\begin{equation*}
   \lambda=\frac{l+1/2}{\sqrt{S}},\,\,\,\nu=\frac{n}{\sqrt{S}},\,\,\,\, \epsilon_{\nu\lambda}=-\frac{1}{S}2m_a r_s^2 E_{nl}.
\end{equation*}
where $0<\lambda,\nu,\epsilon \lesssim 1$.
The WKB wave function of (\ref{scheq}) in the oscillating region, between turning points, $z_{\rm min}$ and $z_{\rm max}$ is
\begin{align}\label{wave}
    \phi_{\nu\lambda}=\frac{\cal{N}}{z F_{\epsilon\lambda} (z)^{1/4}}\cos\left(\sqrt{S}\int_{z_{\rm min}}^z dz'\sqrt{F_{\epsilon\lambda}(z')}\right),
\end{align}
where $\psi_{nl}\sim r_s^{-3/2}\phi_{\nu \lambda}$, and ${\cal N}$ is a normalization factor,
\begin{align*}
    {\cal{N}}=\left(4\pi \int_{z_{\rm min}}^{z_{\rm max}}\frac{ dz}{\sqrt{F_{\epsilon\lambda}}}\right)^{-1/2},
\end{align*}
and where
\begin{align*}
    F_{\epsilon\lambda}(z)=v(z)-\epsilon_{\nu\lambda}-\frac{\lambda^2}{z^2}.
\end{align*}
Notice that we use the indices, $\epsilon$ and $\lambda$, as  labels for $F$ since it explicitly depends on the scaled binding energy, $\epsilon$, and the scaled angular momentum quantum number, $\lambda$. Outside the region of oscillation the wave function, at $z>z_{\rm max}$ and at $z<z_{\rm min}$, drops as
\begin{align}\label{waveexp}
    \phi_{\nu\lambda}\sim \exp\left(-\sqrt{S}\int_{z_{\rm max}}^z dz'\sqrt{-F_{\epsilon\lambda}(z')}\right).
\end{align}The WKB quantization condition is \cite{Langer,Sergeenk0_1996},
\begin{align}\label{quant}
    \int_{z_{\rm min}}^{z_{\rm max}} dz\sqrt{F_{\epsilon\lambda}(z)}=\pi(\nu-\lambda).
\end{align}
Using our choice of $v(0)=1$ and  the solution of the differential equation for $v(z)$, discussed in a subsequent section, we can calculate  $v''(0)=-0.326$. Then the ground state energy (and the energy of other low lying states) can be calculated, using  (\ref{quant}). We obtain
\begin{equation*}
\epsilon_{10}=1-\frac{1.2112}{\sqrt{S}}+O(S^{-1}).
\end{equation*}
The first few low-lying energy levels can be also calculated exactly, and they all are of the form $1-\epsilon_{nl}=O(S^{-1/2}).$

Now, using a collection of data by~\cite{Rodrigues} with a range of virial masses $10^9 M_\odot \lesssim M_\text{vir} \lesssim 10^{13} M_\odot$ and core radii $1 \, \text{kpc} \lesssim r_c \lesssim 10 \, \text{kpc}$, and using the core radius for $r_s$, one can conclude that many galaxies will fall within the physical range of the size parameter $10^2\lesssim S\lesssim 10^6$, so the $S\to\infty$ approximation is appropriate.  The physical range of the radial parameter $z$ is finite. At large $S$, the wave function  (\ref{wave}) oscillates very fast and when taking its square in integrals the square of the trigonometric function in the expression of $\phi_{\nu\lambda}^2$ or $\phi_{\nu\lambda}'^2$ can be replaced by 1/2.  
Furthermore, as $S\to\infty$ the leading order WKB approximation becomes  increasingly reliable, because the size of the transition region between (\ref{wave}) and  (\ref{waveexp}) is $\delta z\sim 1/\sqrt{S}$.
 
\section{Decay of weakly bound states}
Though the contribution of the self-interaction term is neglected here, because we consider the region of $M$ where it is negligible compared to the gravitational term, it cannot be neglected considering an important decay process. Denoting particles in bound states, with quantum numbers $n$ and $l$, and energies $E_{nl}<0$, by $\langle n l|$, and scattering states by $|E_k\rangle$, where $E_k>0$, the self-interaction operator (assuming a standard $\phi^4$ self-interaction for FDM) has nonzero transition matrix elements
\begin{align*}
   {\cal{M}}=\left\langle n_1l_1,n_2l_2\left| \frac{1}{4!} \frac{m_a^2}{f_a^2}\Phi^4\right|n_3l_3,E_k\right\rangle,
\end{align*}
where
\begin{equation*}
    E_k=E_{n_1l_1}+ E_{n_2l_2}- E_{n_3l_3}>0.
\end{equation*}
Then, it is easy to see that every state having energy $E_{nl}>E_{10}/2$, where $E_{10}$ is the ground state energy, is unstable.  On the other hand, states having energy $E_{nl}<E_{10}/2$ are completely stable. The question arises, however, whether sufficient time passes between the creation of the galaxy,  such that all unstable states decay by the time of observation. The detailed investigation of the time dependence of the decay of FDM systems and the formation of the gap is left to a future work~\cite{SSW}.

\section{Solution of the differential equation for the gravitational potential}
Assuming that the energy and particle number of the FDM system are conserved in good approximation, and the decay processes occur sufficiently slowly, through equilibrium states, Bose-Einstein statistics allows the calculation of the occupation number in states labeled by quantum numbers $n$ and $l$. The occupation numbers are enormous, so Bose-Einstein statistics reduces to Rayleigh-Jeans statistics with occupation numbers
\begin{align*}
    N_{nl}=
  \frac{1}{\beta(-\mu+E_{nl})}=\frac{1}{\tilde\beta(-\tilde\mu-\epsilon_{nl})},
\end{align*}
where $\beta$ and $\mu$ are the effective temperature and chemical potential, while $\tilde\beta$ and $\tilde\mu$ are defined as
\begin{align*}
    \tilde\beta=\beta\frac{S}{2 m_a r_s^2},\,\,\,\tilde\mu=\mu\frac{2 m_a r_s^2}{S}.
\end{align*}
Here, $\beta$ and $\mu$ are introduced to ensure energy and particle number conservation, respectively.  As in \cite{Lin}, these are arbitrary model parameters.  However, we will show that the density profile for systems considered here can be scaled to be independent of both $\beta$ and $\mu$. 

 The scaled density function is,
\begin{align*}
   \tilde \rho(z)=\frac{1}{\tilde\beta}\sum_{nl}(2l+1)\frac{1}{-\tilde\mu-\epsilon_{\nu\lambda}}\frac{{\cal N}^2}{z^2\sqrt{F_{\epsilon\lambda}}},
\end{align*}
where the squares of the fast oscillating periodic functions were dropped.

Considering that the energy spectrum, $\epsilon_{\nu\lambda}$, and the values of $\lambda^2$ are dense, summations can be turned into integrations. Then, after changing the integration variable $\nu$ to the variable $\epsilon$, the expression for $\tilde\rho$ takes the form,
\begin{align*}
\tilde  \rho(z)=\frac{1}{\tilde\beta}S^{3/2}\int_{\alpha}^{v(z)} d\epsilon\int d\lambda^2 \left.\frac{d\nu}{d\epsilon}\right|_\lambda\frac{1}{-\tilde\mu-\epsilon}\frac{{\cal N}^2}{z^2\sqrt{F_{\epsilon\lambda}}},
\end{align*}
where we introduced cutoff parameter (gap) $0\leq\alpha<1$ for the scaled energy. Note that the introduction of cutoff parameter $\alpha$ is equivalent to the introduction of virial radius, $r_{\rm vir}$ by $v(r_{\rm vir})/r_s=\alpha$.  The scaling function $v(z)=1/z$ if $r>r_{\rm vir}$. If $\alpha>0$ then FDM is compact. The maximal energy parameter (for the ground state) is close to the maximum of $v(z)$, which is $v(0)=1$. Thus, the FDM system is stable if $\alpha=1/2$, which we consider to be the maximal value. 

Now, using (\ref{quant}), it is easy to establish the relation
\begin{align*}
    {\cal N}^2\left.\frac{d\nu}{d\epsilon}\right|_\lambda=\frac{1}{8\pi^2}.
\end{align*}
Then integrations over variables $\lambda^2$ and $\epsilon$ can be easily computed with the result
\begin{align}\label{density2}
    \tilde\rho(z)=\sigma\left(\sqrt{w(z)}-\sqrt{1-w(z)}\sin^{-1}\sqrt{w(z)}\right),
\end{align}
where
\begin{align}\label{definew}
\sigma=\frac{S^{3/2}}{2\pi^2\tilde\beta}\sqrt{-\tilde\mu-\alpha},\quad w(z)=\frac{v(z)-\alpha}{-\tilde\mu-\alpha}
  \end{align}
  Note that $\tilde\mu>-1$ is unphysical. Otherwise, as shown by (\ref{definew}) we would have $w(0)>1$ and the neighborhood of $z=0$ would not be in the physical domain of (\ref{density2}).

Using definitions (\ref{definew}) and the Poisson equation for the gravitational potential 
\begin{align*}
    \nabla^2v(z)=-4\pi \frac{m_a}{M}\tilde\rho(z),
\end{align*}
we obtain an equation, free of parameters, for $w(x)$, as follows:
\begin{equation}\label{weq}
    \nabla^2w(x)+\sqrt{w(x)}-\sqrt{1-w(x)}\sin^{-1}\sqrt{w(x)}=0,
\end{equation}
where the new scaling variable, $x$, is defined by
\begin{align}\label{relative}
z = x \sqrt{\frac{\pi}{2} \frac{M}{m_a} \frac{\tilde{\beta} \sqrt{-\tilde{\mu} - \alpha}}{S^{3/2}}}.
\end{align}

\begin{figure}[b]
\includegraphics{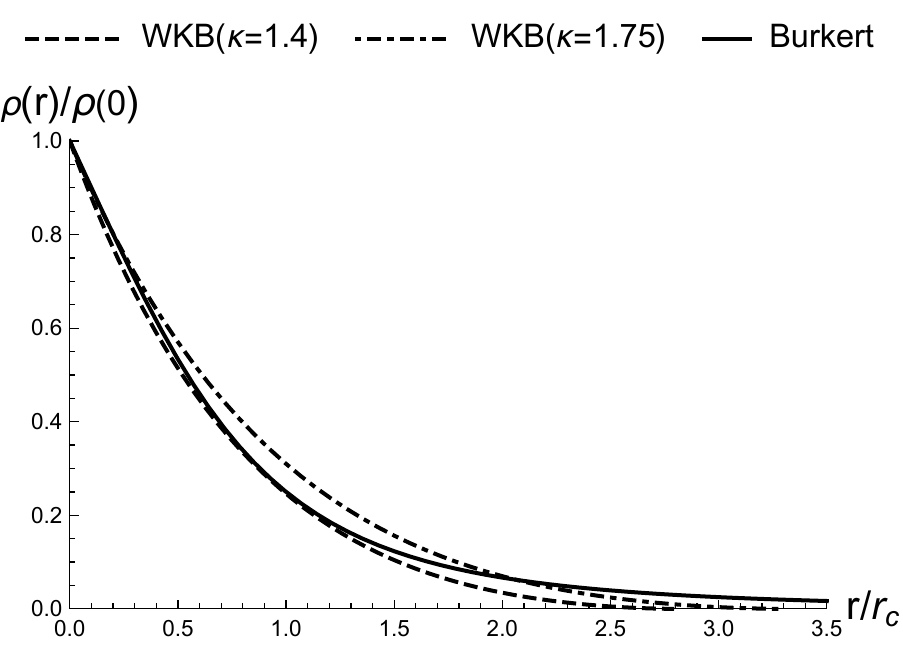}
\caption{\label{density}Comparing  profile (\ref{master}), rescaled by factor $\kappa=r_s/r_c$, at $\kappa=1.4$, $\kappa=1,75$, and the Burkert profile (\ref{bur})
. }
\end{figure}

Then the central result of this letter is the following expression for the profile of FDM:
\begin{align}\label{master}
    \rho(r)=\rho_c \left(\sqrt{w(r/r_c)}-\sqrt{1-w(r/r_c)}\sin^{-1}\sqrt{w(r/r_c)}\right),
\end{align}
where $\rho_c$ is the core density and $r_c$ is the core radius.
The properties of the scaling function,  $\rho$, will be discussed in the next section.
\section{The profile of fuzzy dark matter}\label{sec:FDM_profile}
We chose integration constants $w(0)$ and $w'(0)$ for solving (\ref{weq}). Notice that the solution of  the equation  is real only if $w(0)$ satisfies the constraint $0<w(0)\leq 1$.  No matter our choice for $w(0)$, we must choose the second integration constant as $w'(0)=0$, otherwise, due to the singularity of the Laplacian $w''(x)+2 w'(x)/x$ at $x=0$, the solution for $w(x)$ is singular.  Consequently, the single parameter $w(0)$ defines a unique solution.  Fixing $w(0)$ determines the chemical potential $\tilde\mu$, as well, because using $v(0)=1$, (\ref{definew}) implies
\begin{equation*}
\tilde\mu=-\alpha\left(1+\frac{1}{w(0)}\right)\leq -2\alpha.
\end{equation*}
Then unless $1-w(0)\ll 1$  all occupancies are of similar magnitude,
\begin{equation*}
N_{nl}= O\left(\frac{1}{\tilde\beta\, \tilde\mu}\right).
\end{equation*}
The alternative is that $1-w(0)\ll 1$, implying $-\tilde\mu-2\alpha\ll1$, as well.  Then low lying states, along with  the ground state, have occupation numbers much larger than other stable states, with smaller binding energy parameters.

There is reason to believe that the second alternative is realized in FDM systems.  No matter what chemical potential is reached in the collapse forming FDM, the decay process strongly skews the distribution of occupancy numbers towards deeply bound states. Consider that in a decay process two axions in unstable states, with binding energy parameters $\alpha<\epsilon_1<1/2$ and $\alpha<\epsilon_2<1/2$ are annihilated while a stable axion with $1>\epsilon_3>1/2$ and a scattering state axion are created.  The question is how the transition probability depends on $\epsilon_3$. There are two reasons why transitions into stable states with larger binding energy have higher probability: (1) the  number of decay channels, or in other words, the range of unstable states that can take part in the decay process, $\epsilon_1+\epsilon_2 <\epsilon_3$  increases with increasing $\epsilon_3$, and (2) the transition probability is also proportional to the phase space of the emitted scattering state axion, which, in turn, is proportional to $\epsilon_3-\epsilon_1-\epsilon_2$, which also increases with $\epsilon_3$.  The net result is that the occupancy number of a stable state becomes an increasing function of its binding energy.   If the decay process proceeds slow enough, then the system of stable states can be considered to  be in near equilibrium during the process.  Then the decays will lead to a monotonic increase of $\tilde\mu$, as the ratio of occupancies of states of different energies depends only on $\tilde\mu$.  Considering the upper bound
$\tilde\mu\leq-2\alpha,$ it is reasonable to expect that $-\tilde\mu-2\alpha\ll1$, which will be the main focus of further discussions.  However, we will also consider briefly the case when $-\tilde\mu-2\alpha=O(1)$.
\begin{figure}[b]
\includegraphics[width=0.5\textwidth]{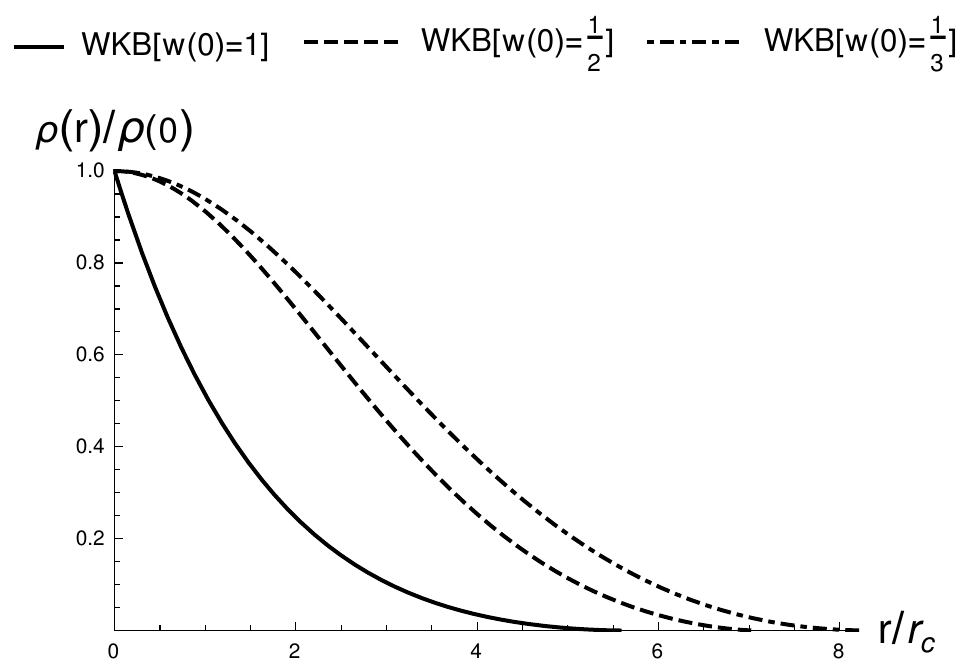}
\caption{\label{v0values}Comparing  profiles (\ref{master})for initial conditions $w(0)=1$ ($\tilde\mu=-1$), $w(0)=1/2$ ($\tilde\mu=-2+\alpha$), and $w(0)=1/3$ ($\tilde\mu=-3+2\alpha$).}
\end{figure}

One may wonder how the effective temperature, $\tilde\beta$, varies during the decay process, in which the total mass of stable states and $\tilde\mu$ both increase. Since the mass is expected to increase slowly, as its total possible increase is just half of the total mass of unstable states, it is likely that $\tilde\beta$ is forced to increase, as well.  Then the decay process is just a quantum evaporation cooling process, similar to those in atomic physics, creating Bose-Einstein condensates. The difference is that FDM systems considered here contain a lot of stable excited states, in addition to the condensate. The time dependence of the decay process will be investigated in a forthcoming publication~\cite{SSW}.

Before investigating the full solution of (\ref{weq}), we consider the behavior of the solution near two endpoints of the distribution.  Expanding (\ref{weq}) around $x=0$ we obtain
\begin{equation}\label{w2prime}
w''(0)=-\frac{1}{6}\left(\sqrt{w(0)}-\sqrt{1-w(0)}\sin^{-1}(\sqrt{w(0})\right).
\end{equation}
(\ref{w2prime}) can be used to calculate the binding energy and radial dimensions (the difference between the turning points) of the ground state and other low lying states. For those states, the higher derivatives of $w$ can be neglected as they are suppressed by orders of $S^{-1/2}$. 

Consider now that $v(z)$ decreases monotonically from $v(0)=1$ to $v(z_\text{vir})=\alpha$. Then, it follows from (\ref{definew}) and (\ref{density2}) that $w(z_\text{vir})=0$ and $\tilde\rho(z_{\rm vir})=0$. Expanding $\rho(z)$ with respect to $w(z)$ around $w(z) = 0$ in (\ref{density2}) we obtain
\begin{equation*}
\tilde\rho(z)\simeq\frac{\sigma}{3}w(z)^{3/2}\sim (z_\text{vir}-z)^{3/2},
\end{equation*}
implying that the profile of FDM is compact. By necessity, a cutoff was also imposed on the energy distribution  in~\cite{Lin}.  Consider now that at $z>z_\text{vir}$ gravity is Newtonian, $v(z)=1/z$, Then  continuity of $v(z)$ and $v(z_\text{vir})=\alpha$ implies that $z_{\rm vir}=1/\alpha$.

There is no solution when $\alpha=0$, corresponding to infinite virial radius.  To see that consider that (\ref{weq}) has a unique solution for every choice of the initial condition, $w(0)$, which has a finite range, vanishing at a finite value, $x=x_{\rm vir}$. Then it follows from (\ref{relative})  that $z_{\rm vir}=x_{\rm vir}/\alpha$, implying that  the scaling factor between $x$ and $z$ is given by $z=x/\alpha\to\infty$. The only possible solution is then $w(z)\equiv0$, which is the solution of (\ref{weq}) with initial conditions $w(0)=0$.

There is no analytic solution of (\ref{weq}) in terms of known transcendental functions.  However, numerical solutions for every choice of $w(0)$ are easily obtained.  Choosing $w(0)=1$, corresponding to $\tilde\mu=-\alpha$, the numerical solution vanishes at  $
x_\text{vir}=5.536$, showing that the physical range of $x$ is $0\leq x\leq 5.536,$  corresponding to the physical range of $0\leq z\leq 1/\alpha$ .  This provides the scaling factor of  $x/z=\alpha  5.536$. Decreasing $w(0)$  increases $
x_\text{vir}$, as discussed later.

Comparing profile (\ref{master}) with phenomenological profiles of dark matter, obtained from fitting to rotation curves of stars, we must consider that the scaling factor, $r_s$, is arbitrary.  Since our profile does not have a singular cusp near the origin, comparison with the Navarro-Frenk-White (NFW) profile~\cite{NFW} at low values of $z$ is difficult. However, comparison with the phenomenological Burkert profile~\cite{Burkert}, which has a finite cusp only,
\begin{equation}\label{bur}
\rho(z)=\frac{\rho(0)}{\left[1+\frac{r}{r_c}\right]\left[1+\left(\frac{r}{r_c}\right)^2\right]},
\end{equation}
 is possible. Observe now that profile (\ref{bur}) is associated with a specific definition of the scaling radius, called core radius, $r_c$. Our scaling radius, $r_s$, is defined  in (\ref{rsdef}).  Using the Burkert profile (\ref{bur}), the core radius, $r_c$, does not satisfy (\ref{rsdef}).  Therefore, to compare the two profiles we must introduce an extra rescaling factor. Using two different rescaling factors $\kappa \equiv r_s/r_c = 1.4$ and $1.75$ we compare profiles  (\ref{master}) and (\ref{bur}) in Fig. \ref{density}.

Just like the NFW profile, the Burkert profile also requires a finite cutoff, so that the integral for the expression of the total mass would be finite. The choices for  rescaling factor $\kappa = 1.4$ and $1.75$ correspond to setting the virial radius of the Burkert profile at $z_\text{vir} \sim 4$.

The NFW profile, having a cusp at the origin, cannot be fitted very well. We note that adding effects due to a central black hole and standard matter will change the gravitational potential and, subsequently, the density profile (\ref{master}).  We leave such an analysis for future studies.

Finally, we will investigate solutions for initial condition $w(0)<1$.  Numerical solutions for $w(x)$ are obtained again solving (\ref{weq}), while Fig. \ref{v0values} shows the density profile for three different values of $\tilde{\mu}$.  Note that  $\rho'(0)=0$ for solutions with initial conditions $v(0)< 1$. Data for most galaxies do not have such profiles, indicating that in general $0<-\tilde\mu-1\ll 1$.

\section{Summary}
Assuming that galactic dark matter is fuzzy dark matter in excited states of the equations of motion,  the equations can be solved using the WKB approximation. The approximation improves with the increase of the size of the galaxy, but it is expected that the leading order WKB approximation gives reasonably good energy eigenvalues and eigenfunctions even for smaller galaxies. 

Axions are expected to have four-particle self-interactions, $L_{\rm int}\sim \lambda \Phi^4$ as part of an instanton potential~\cite{Hui}. In the Born approximation of the self-interaction potential, excited states with binding energy $E< E_{10}/2$, where $E_{10}$ is the binding energy of the ground state, decay in an energy conserving process, which is possibly similar to quantum radiation cooling. The decay is accompanied by the emission of scattering state particles from the galaxy, generating a gap in the energy spectrum. As a result, after, or possibly simultanously with the creation of the galaxy, the energy spectrum cuts off at half of the binding energy of the ground state. The relative timing of the collapse of the overdensity of axions into fuzzy dark matter and of the decay process is yet to be determined. The emitted particles may serve as seeds of future galaxies. 

The WKB method has been used to calculate approximate wave functions (in the large $S$ limit), as analytic functions of the gravitational potential.  The density function of dark matter was built from the WKB wave-functions, using Bose-Einstein statistics at equilibrium. Combined with the Poisson equation, a differential equation is found for the rescaled gravitational potential. Though there are no undetermined constants in the equation, one integration constant, restricted to a finite physical range must be fixed.  The decay process likely drives the constant close to the end of the physical range, resulting in a well determined density profile for fuzzy dark matter, which contains only the usual two free parameters, the central density and the core radius.  Having an energy gap results in a well-defined cutoff of the density profile.  With appropriately chosen scale parameters the density distribution is close to that of the Burkert profile~\cite{Burkert}.

\section{acknowledgments}
The authors are indebted to Joshua Eby for fruitful discussions.  L.S. and L.C.R.W. thank the University of Cincinnati Office of Research Faculty Bridge Program for funding through the Faculty Bridge Grant.  L.S. also thanks the Department of Physics at the University of Cincinnati for financial support in the form of the Violet M. Diller Fellowship.


\begin{thebibliography}{100}
\bibitem{footnote1} Note that a cutoff in the energy spectrum was introduced in~\cite{Lin}, as well, but that self-interactions introduce such a cutoff in a natural way.
\end{thebibliography}
\end{document}